\documentclass[sigchi]{acmart}

\AtBeginDocument{%
  \providecommand\BibTeX{{%
    \normalfont B\kern-0.5em{\scshape i\kern-0.25em b}\kern-0.8em\TeX}}}

\setcopyright{acmcopyright}
\copyrightyear{2018}
\acmYear{2018}
\acmDOI{10.1145/1122445.1122456}

\acmConference[CIKM 2019]{CIKM 2019 }{Novembre 03--07, 2018}{Beijing, CN}

\usepackage[english]{babel}
\usepackage[utf8]{inputenc}
\usepackage{libertine}
\usepackage{graphicx}
\usepackage{floatflt}
\usepackage{blindtext}
\usepackage{enumitem}
\usepackage{amsthm}
\usepackage{subfig}
\usepackage{listings}
\usepackage{amsmath}
\usepackage{framed}
\usepackage{minibox}
\usepackage{float}
\usepackage{wrapfig}
\usepackage{longtable}
\usepackage[strict]{changepage}

\usepackage{xcolor} 
\definecolor{light-gray}{gray}{0.95}
\definecolor{mygreen}{rgb}{0,0.6,0}
\definecolor{mymauve}{rgb}{0.58,0,0.82}

\usepackage{listings}

\begin{document}

\title{An interactive dashboard for searching and comparing soccer performance scores}
\author{Paolo Cintia}
\authornotemark[1]
\email{paolo.cintia@di.unipi.it}
\orcid{}
\affiliation{%
  \institution{ University of Pisa, Italy}
  \streetaddress{Largo B. Pontecorvo 3}
}

\author{Giovanni Mauro}
\email{g.mauro7@studenti.unipi.it}
\affiliation{%
  \institution{ University of Pisa, Italy}
  \streetaddress{Largo B. Pontecorvo 3}
}

\author{Luca Pappalardo}
\authornotemark[1]
\email{luca.pappalardo@isti.cnr.it}
\affiliation{%
  \institution{ISTI-CNR, Italy}
  \streetaddress{Via G. Moruzzi 1}
  }

\author{Paolo Ferragina}
\email{paolo.ferragina@unipi.it}
\affiliation{%
  \institution{ University of Pisa, Italy}
}

\renewcommand{\shortauthors}{Cintia et al.}

\begin{abstract}
The performance of soccer players is one of most discussed aspects by many actors in the soccer industry: from supporters to journalists, from coaches to talent scouts. 
Unfortunately, the dashboards available online provide no effective way to compare the evolution of the performance of players or to find players behaving similarly on the field. This paper describes the design of a web dashboard that interacts via APIs with a performance evaluation algorithm and provides graphical tools that allow the user to perform many tasks, such as to search or compare players by age, role or trend of growth in their performance, find similar players based on their pitching behavior, change the algorithm's parameters to obtain customized performance scores. We also describe an example of how a talent scout can interact with the dashboard to find young, promising talents. 
\end{abstract}



\maketitle

\section{Introduction}
\label{sec:Introduction}
Nowadays many actors in the soccer industry, from television broadcasters to scouts and professionals in soccer clubs, are increasingly relying on data-driven scores to rank players, find promising talents, and increase fan engagement \cite{gudmundsson2017spatio, pappalardo2017quantifying, bornn2018soccer}. 
Several online platforms, such as wyscout.com or whoscored.com,  allow for searching players in a database and show aggregated statistics of their performance. Unfortunately, these tools provide no intuitive way to \emph{compare} the \emph{evolution} of performance of players or \emph{suggest} players behaving in a similar manner. 

In this paper we describe a web dashboard for searching and comparing data-driven performance scores of soccer players. 
The dashboard provides the user with a graphical interface to interact with the {\sf PlayeRank} algorithm \cite{pappalardo2018playerank} that offers a principled evaluation of the performance of players based on data describing all the spatio-temporal events (e.g., passes, shots, etc.) that occur during a match.
Several actors in the sports industry may benefit from our dashboard:
\begin{itemize} 
    \item a \emph{talent scout}, who searches for promising talents that meet specific constraints (e.g., age or role);
    \item a \emph{coach}, who needs to visualize the evolution of the performance of their players to select the team's lineup in the next match;
    \item a \emph{sports journalist}, who wants to comment in an article about the performances in a recent match;
    \item a \emph{soccer enthusiast}, who wants a support to set up the lineup of their fantasy football team.
\end{itemize}

The design of the dashboard is hence motivated by the need of providing these actors with: (i) a way to visualize the evolution of the performance of a player in time; (ii) a compact way to compare the performance of two or more players; (iii) the possibility to search players by role and to filter them according to specific constraints (e.g., age, trend of growth of their performance, matches played); (iv) the possibility to change the parameters of the  {\sf PlayeRank} algorithm to obtain tailored evaluations of performance. 
A demo-video of the web dashboard is available at the following link: \url{youtu.be/SzrDRKucRjE}, while an online version will be available soon at \url{playerank.d4science.org}.

\section{Dashboard Architecture}
\label{sec:Architecture}

\begin{figure}[htb]
	\centering\includegraphics[scale=0.3]{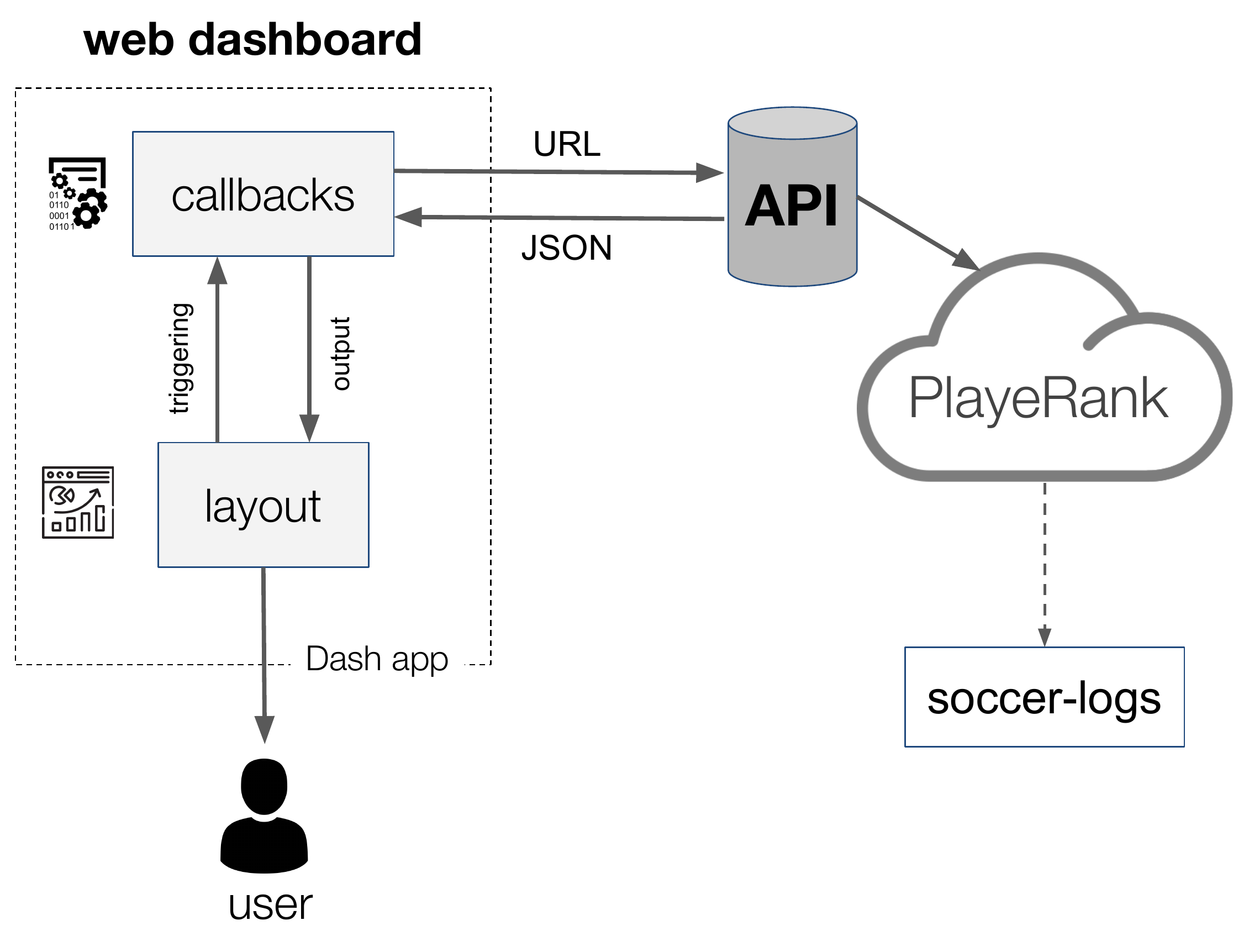}
	\caption{Schema of the communication between the web dashboard and the API. 
	}
	\label{fig:schema1}
\end{figure}

The web dashboard is designed in Python as a Dash \cite{Dash} app that communicates with the {\sf PlayeRank} algorithm \cite{pappalardo2018playerank} (\autoref{fig:schema1}). The communication channel is realized through an API that implements the exchange of data in two directions: (i)
 the web dashboard sends a request for aggregated data through the HTTP protocol, using an URL containing the parameters of the request; 
 (ii) the API returns the desired aggregated data using JSON format.

A Dash app consists of two parts:  the layout and the set of callbacks (\autoref{fig:schema1}).
In the layout, all the components of the graphical interface are specified, each having a unique and unambiguous identifier. This identifier is the connection point to the callbacks part, a dynamic section that specifies the actions to be
executed when an event occurs on a layout component (e.g., the user clicks on a button or
changes the value in a dropdown). The callbacks are the part that actually communicates with the API functions. This communication is transparent to the user, who can only interact with the layout part of the dashboard.

\section{Dashboard Layout}
\label{sec:Platform}

\begin{figure}[htb]
\centering
\includegraphics[scale = 0.35]{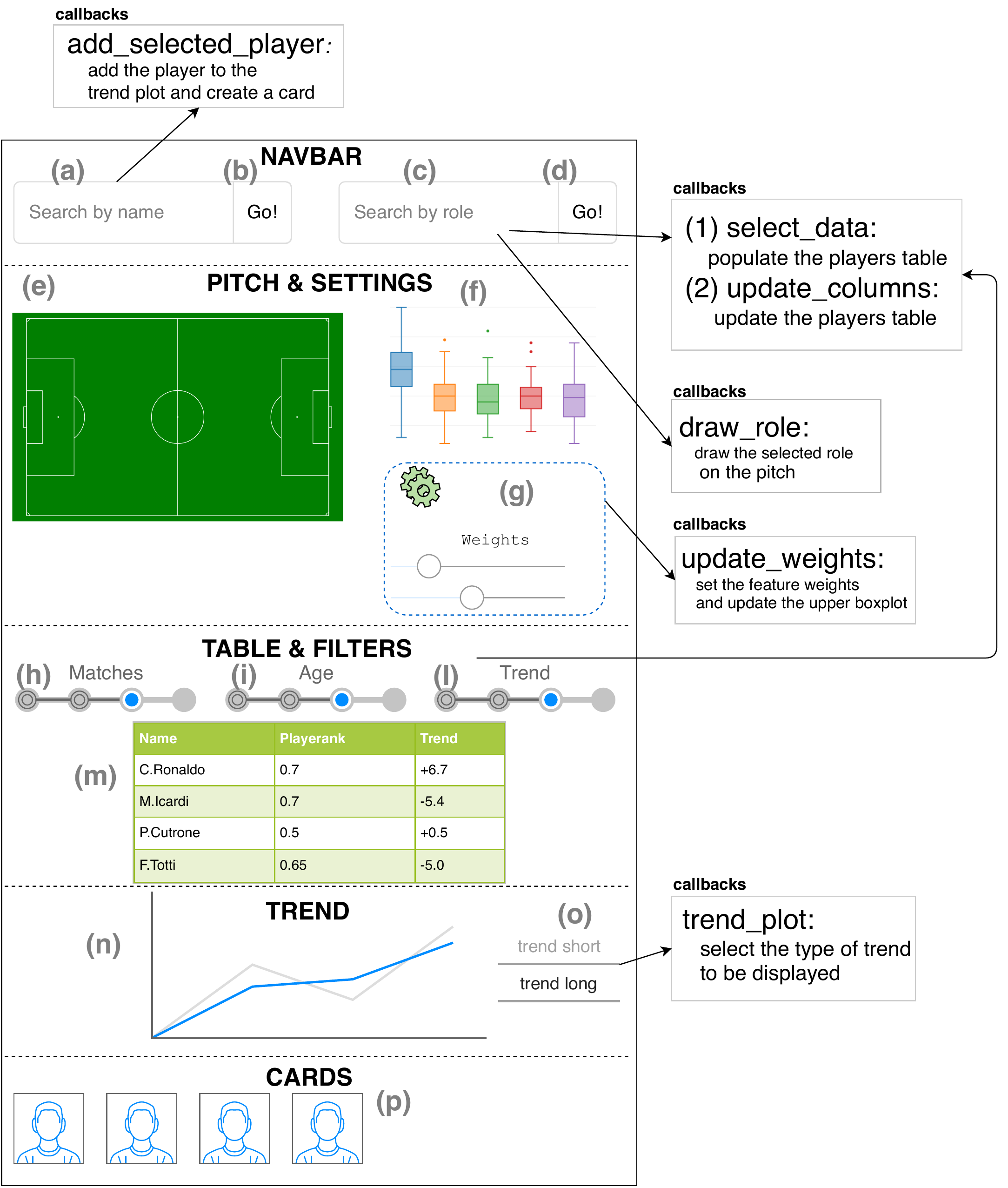}
\caption{Graphical components in the dashboard's layout. The boxes list the callbacks that are triggered when the user interacts with a layout component. } 
\label{fig:Scheme}
\end{figure}

Figure \ref{fig:Scheme} shows the components in the layout of the web dashboard and the callbacks associated with each graphical component. The design of the layout has been guided by the \emph{Visual Exploration Paradigm} \cite{du2010visual}, consisting in a three-step process, called the \emph{Information Seeking Mantra}:  overview, zoom and filter, and details-on-demand.  

The upper part of the layout (Navbar) is a navigation bar containing two search dropdowns (\autoref{fig:Scheme}a, c) with their corresponding buttons (\autoref{fig:Scheme}b, d). The dropdowns are associated with callbacks that: (i) call the API to retrieve all players having the name or role typed by the user (\autoref{fig:Scheme}); (ii) highlight on the pitch part the selected role; (iii) add the selected players in the players table.

The second part of the layout (Pitch \& Settings) contains two elements: the visualization of a soccer pitch that highlights the roles selected in the navigation bar; and the settings panel (\autoref{fig:Scheme}g) that, when its filters are modified, updates the boxplot visualizing the distribution of the players' performance score per role (\autoref{fig:Scheme}f). 

The third part (Table \& Settings) contains a set of sliders (\autoref{fig:Scheme}h, i, l) that allow the user to filter the players by age, trend of performance growth and number of matches played. Sliders are associated with a callback that updates the players table, a component that visualizes the names of the selected players, as well as other information like age and role, average performance score and trend of performance growth (\autoref{fig:Scheme}m). 

The fourth part (Trend) contains a line chart (\autoref{fig:Scheme}n) visualizing the evolution of the performance scores of the players that have been selected in the players table. 

Finally, the last part (Cards) contains some cards that show further information about the selected players and, in addition, it shows other players that are {\em similar} according to the way they have pitched on the soccer field (\autoref{fig:Scheme}p). 


In the next sections we describe in detail some of the layout components of the web dashboard, in order to explore its functionalities, and refer the interested reader to the demo-video available at the following link: \url{youtu.be/SzrDRKucRjE}. 

\subsection{Pitch Plot}

\begin{figure}[htb]
\centering
\includegraphics[scale = 0.4]{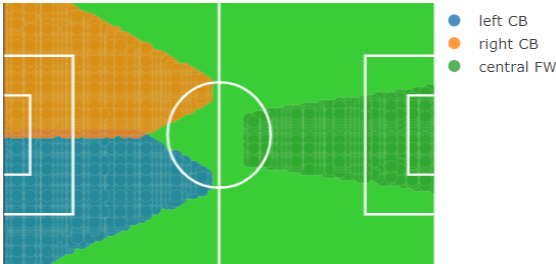}
\caption{Drawing of three roles on the soccer pitch plot: left central back (left CB), right central back (right CB) and central forward (central FW).}
\label{fig:Pitch}
\end{figure}

The soccer pitch component (\autoref{fig:Pitch}) is linked to the ``Search by role'' dropdown in the navigation bar. 
When the value of the dropdown changes, a callback is triggered, drawing the corresponding role data on the soccer pitch component. The role is drawn by highlighting the positions of the pitch associated with that role. \autoref{fig:Pitch} shows how the soccer pitch components looks like when three roles (left CB, right CB and central FW) are selected in the ``Search by role'' dropdown.

\subsection{Settings panel}
\label{sec:settings}
\begin{figure}[htb]
\centering
\includegraphics[scale=0.4]{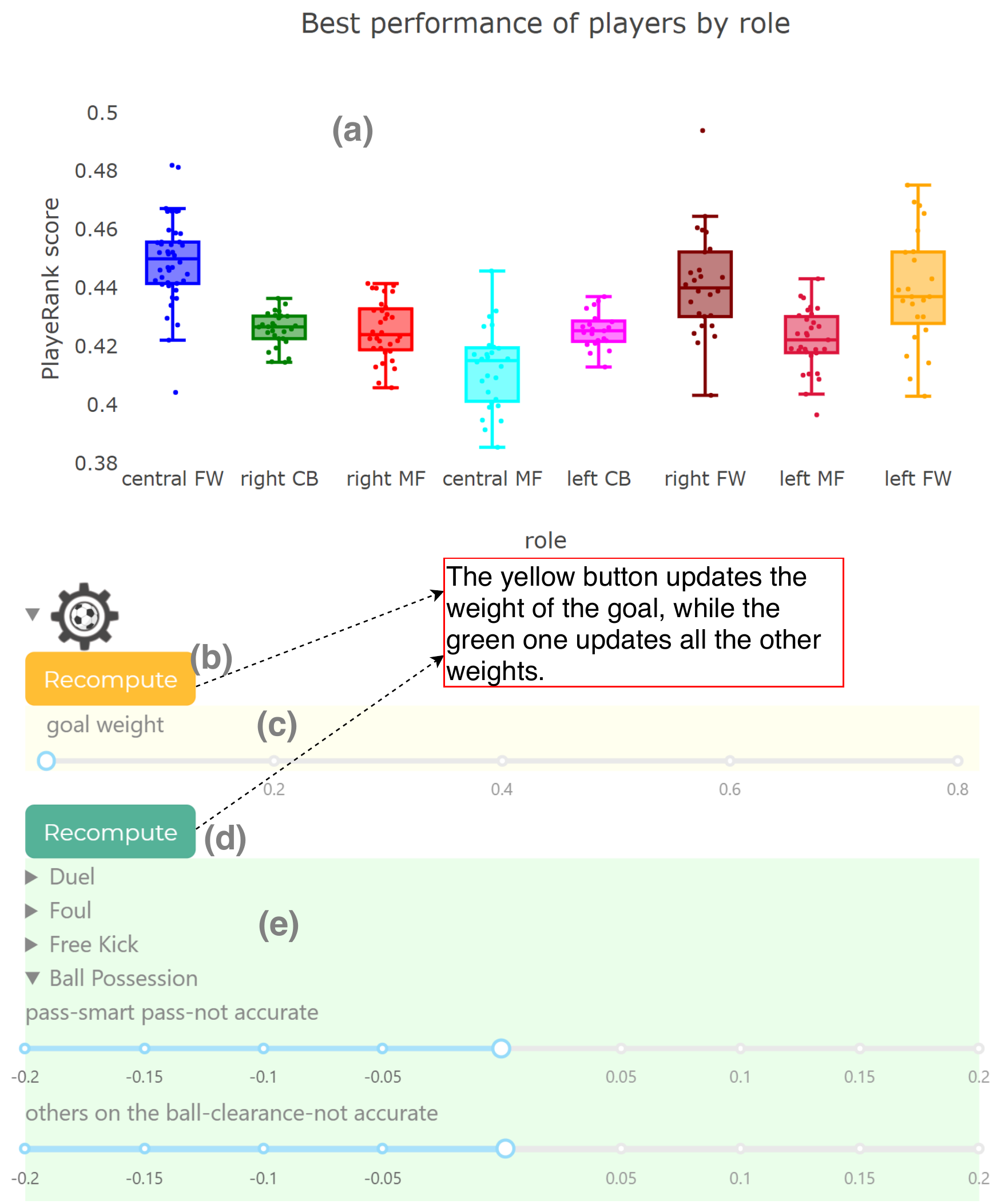}
\caption{Illustration of the Settings panel}
\label{fig:Boxplot}
\end{figure}

In general, {\sf PlayeRank} \cite{pappalardo2018playerank} computes a player's performance score in a match as a scalar product between a vector of features describing their performance (e.g., number of shots, number of cards, expected goals, etc.) and a vector of weights specifying the important of each feature. The web dashboard allows the user to recalibrate the feature weights of the {\sf PlayeRank} algorithm so as to obtain tailored evaluations of performance. 

As an instance, the user can change the importance of scoring a goal into the player evaluation (using the slider in Figure \ref{fig:Boxplot}c), so to reward more the players who score goals. Similarly, the weight associated with each performance feature can be changed by the corresponding slider (Figure \ref{fig:Boxplot}e), which triggers a callback that in turn asks the API to recompute the performance scores with the new weights. 

The boxplot in Figure \ref{fig:Boxplot}a provides a visual summary of the distribution of performance scores per role.
It is updated every time a feature weight is changed by the corresponding slider, i.e., the two buttons in  \autoref{fig:Boxplot}b-d activate a callback that re-draw the boxplot with the new performance scores.
Note that changing the feature weights from the Settings panel implies changing the visualization of the performance scores in the trend plot (\autoref{fig:trends}) as well.

\subsection{Trend plot}
\label{sec:Selected}

\begin{figure*}[htb]
	\centering\includegraphics[scale=0.455]{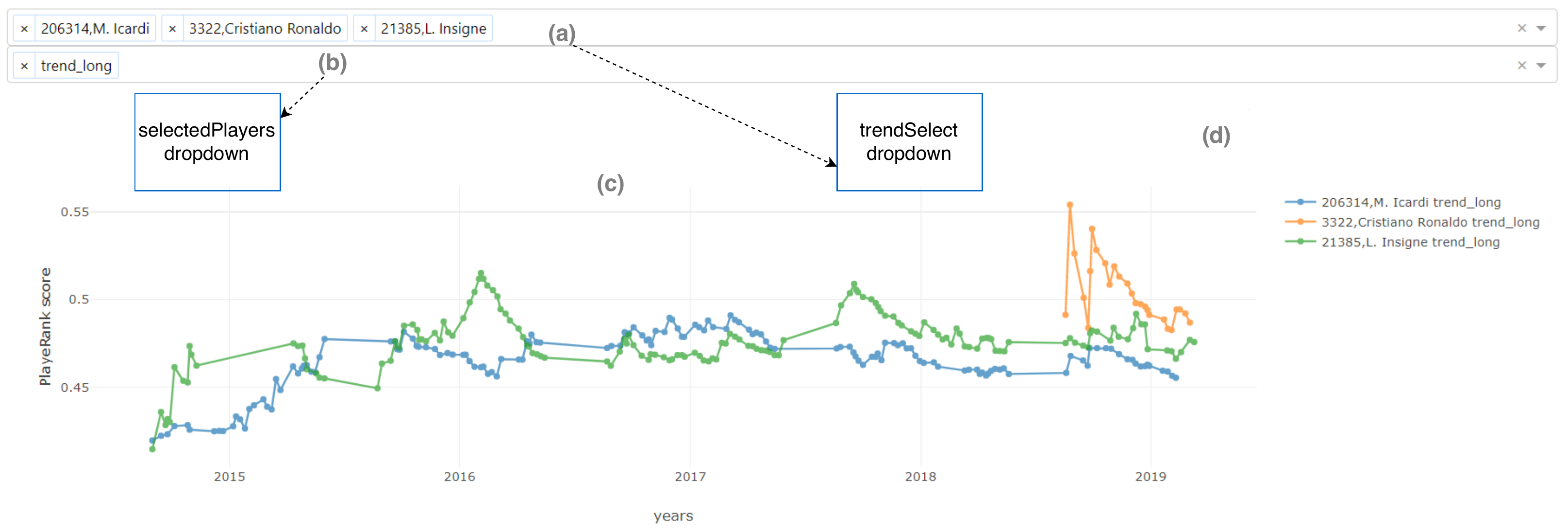}
	\caption{Comparing the performance scores of Icardi, Ronaldo and Insigne (blue, orange and green curves, respectively).}
	\label{fig:trends}
\end{figure*}

The trend plot (\autoref{fig:trends}) allows the user to compare the  evolution in time of the performance scores of soccer players based on {\sf PlayeRank}. For instance, \autoref{fig:trends} compares three top players in the Italian first division season 2018/2019: Mauro Icardi (FC Internazionale), Cristiano Ronaldo (Juventus FC) and Lorenzo Insigne (SSC Napoli). The striking impact of Cristiano Ronaldo (the orange curve) on the Italian league is immediately recognizable from the plot: after a shaky start, probably due to adaptation to a new league, the performance scores of C. Ronaldo shortly overtake the ones of renowned strikers in the league, like Icardi and Insigne.

By using a proper dropdown (\autoref{fig:trends}b) the user can choose among two types of trends: (i) \emph{trend\_long}, calculated taking equally into account all matches' scores; (ii) \emph{trend\_short} which weights more the player's score in the most recent matches.
The trends specified by the user in the dropdown are drawn, via a dedicated callback and for each player, in the trend plot.

\section{The talent scout use scenario}
To demonstrate the usefulness of our web dashboard, we consider a crucial task in the soccer industry such as talent scouting. The main purpose of a talent scout is searching for promising talents, i.e., \emph{young} and \emph{unknown} players who show a good and a positive trend in their performance. A scout can achieve this goal using the web dashboard as follows.

\begin{figure}[htb]
\centering\includegraphics[scale=0.34]{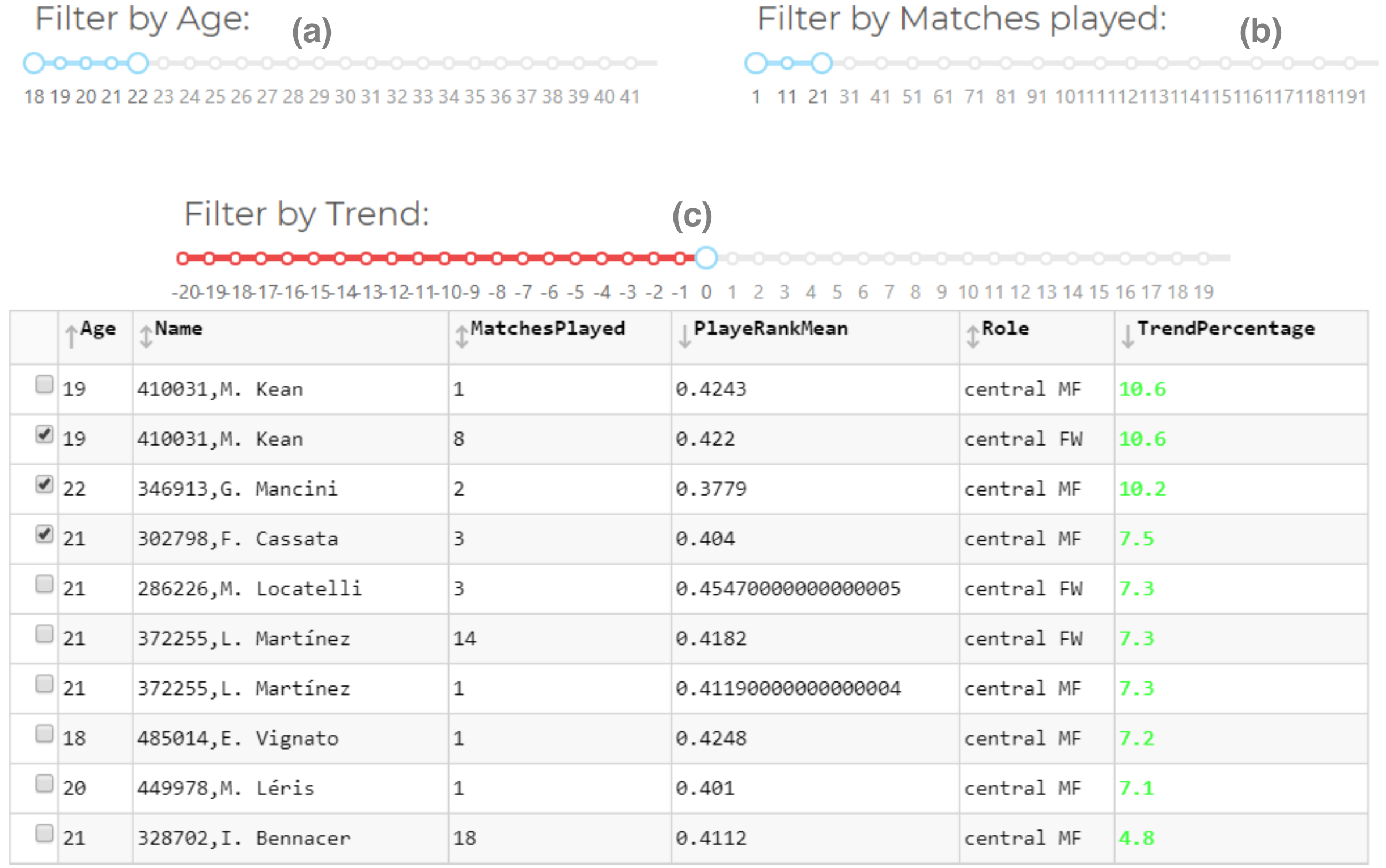}	\caption{Players table resulting from the interaction of the talent scout with the web dashboard.}
	\label{fig:talent}
\end{figure}

The scout first searches for players of specific roles (e.g., central forward or central midfielder) using the ``Search by role'' dropdown. Then, s/he uses the age slider (\autoref{fig:talent}a) to select young players (< 22 yo). To focus on players who promise a bright future, the scout selects a positive trend of growth using the trend slider (\autoref{fig:talent}c).  
The scout sorts the resulting players by trend of growth (column \texttt{TrendPercentage} in \autoref{fig:talent}), by age and finally by average performance score (column \texttt{PlayeRankMean}).

\autoref{fig:talent} shows the players table resulting from the above described operations.
Note that a player can appear in several rows because he can play different roles in different matches. For example Kean occurs twice in the ordered list, once as central MF and once as central FW.

The scout can eventually select some players in the table by using the appropriate check boxes, hence visualizing the evolution of their performance in a specific trend plot. Referring to the example shown in \autoref{fig:talent}, the talent scout selects Kean, Mancini and Cassata as the most promising young talents in the Italian first division and asks the interface to show their performance plot. 

\section{Conclusions}

In this paper we presented a web dashboard for searching and comparing soccer performance scores. Through a set of API endpoints, the web dashboard can retrieve data about performances and players.
Users can search for players by name or roles, getting an immediate comparison of selected players and their performance evolution. Users can also search for players behaving similarly in the previous played games. If we consider that player scouting in soccer is a high-consuming task, our dashboard can help scouts in saving a considerable amount of time, hence partially automating the complex process of finding promising talents.

\begin{acks}
This work has been partially funded by EU project SoBigData RI, grant \#654024. 
\end{acks}

\bibliographystyle{ACM-Reference-Format}
\bibliography{references}

\end{document}